\documentstyle[prl,aps,epsf,psfig,multicol]{revtex}
\begin{document}
\draft 
\title{Stability   of Ge-related 
point defects and complexes in Ge-doped SiO$_2$}   
\author{Carlo Maria Carbonaro, Vincenzo  Fiorentini, and
Fabio Bernardini}
\address{INFM and Dipartimento di  Fisica, Universit\`a di Cagliari,
Cittadella  Universitaria, I-09042 Monserrato (CA), Italy}
\date{Submitted to Phys. Rev. Lett. Jan 10, 2002} 
\maketitle

\begin{abstract}
We analyze  Ge-related defects in Ge-doped
SiO$_2$ using first-principles density functional techniques.
Ge is incorporated at the level of $\sim$1 mol \% and
above. The  growth conditions of Ge:SiO$_2$ naturally
set up oxygen deficiency, with vacancy concentration increasing
by a factor  10$^5$ over undoped SiO$_2$, and O vacancies binding
 strongly to Ge impurities. All the centers considered 
  exhibit potentially EPR-active states, candidates for the identification
of the Ge($n$) centers. Substitutional Ge produces an apparent gap
shrinking via its extrinsic levels.
\end{abstract}
  % point def.

\begin{multicols}{2}
Silicon dioxide, besides its role in silicon-based microelectronics,
is the central material of fiber optics technology.
Germanium doping of  silicon dioxide is a key to a number of technologically
relevant applications, as e.g. low-loss and modulated-refraction fibers.
Ge can be introduced in SiO$_2$ in
concentrations  in the 1--10 \% range \cite{FGS}.
As is generally true of defects in solids, the
identification and characterization of Ge-related  defect centers  
in SiO$_2$ is a difficult task,  and far from completion despite 
intense recent investigation. This is the case, for instance,
for  the understanding of oxygen-deficient Ge-related centers
and the  so called Ge($n$) [$n$=0,1,2,3] centers,
for which rather disparate models have been proposed.
First-principles defects theory  \cite{noi,blo,Pacch} can play a key
role in this context, predicting the stability regimes and
concentrations of Ge and related defects in SiO$_2$, and their
extrinsic electronic levels and potential magnetically active states.
Recently, for example,  these techniques were used  to pinpoint the
correlation of the E$'$ center in  SiO$_2$ with the singly-charged O
vacancy \cite{noi} and the role of hydrogen in determining the leakage
current across thin silica layers \cite{blo}.

  Using a similar methodology,  here
we study the energetics,  extrinsic levels, and solubility of Ge in 
SiO$_2$. We predict experimental signatures of
a selection of Ge-related  defects, including Ge-oxygen vacancy complexes
in various stable and metastable configurations and
charge states. We demonstrate that  oxygen
off-stoichiometry occurs naturally in  Ge-doped  SiO$_2$, and 
specifically that oxygen deficiency (i.e. O vacancy formation)  occurs
preferentially  near substitutional Ge$_{\rm Si}$   sites. 
The electronic structure of  Ge$_{\rm Si}$  may
explain the observed effective reduction of the gap in Ge-doped
silica    \cite{carb}.  Single and paired substitutional Ge
 and their complexes with an O vacancy are all found to have
accessible paramagnetic states:
Ge$_{\rm Si}$, and the Ge$_{\rm Si}$-Ge$_{\rm Si}$  or
Ge$_{\rm Si}$-V$_{\rm O}$-Ge$_{\rm Si}$ complexes, are candidates, 
respectively, for the Ge(1) and Ge(2) electron-capturing centers;
 two metastable E$'$-like Ge-related centers correlate with
the  Ge(3)  hole center.

Our method  can be summarized as follows. The equilibrium
 concentration of a defect  D$= {\rm N}_s \exp 
(-{\rm F}_{\rm form}/{\rm k}_{\rm B} {\rm T}_{\rm g})$
is determined by the growth temperature T$_{\rm g}$, the  number 
 N$_s$  of   available sites, and the formation free  energy 
F$_{\rm form}$=E$_{\rm form}$--T\,S$_{\rm form}$. The latter 
depends \cite{7} on 
 the  chemical potentials of atoms added or removed, on the defect
 charge state, i.e. the charge  released to or captured from the
   electrons reservoir made up by the whole crystal. Given the  
 formation energies of the relevant defects, 
 the   concentrations and the electron  chemical potential $\mu_e$
 are determined  self-consistently to satisfy charge neutrality, as
 detailed in  \cite{7}. A specific  defect configuration or charge
 state is predicted to exist if its  formation energy is lower than
 that of all  other defect states for some value of $\mu_e$, or when 
a sufficiently high energy barrier prevents its disappearance. The
 formation energy  for a defect in charge state $Q$ can be written
 as 
\begin{eqnarray}
{\rm E}_{\rm form}^Q = {\rm E}_{\rm def}^Q
 + Q \mu_e + {\rm M}(Q) 
 - {\rm n}_{\rm Si}\mu_{\rm Si} -{\rm n}_{\rm Ge}\mu_{\rm Ge}
 - {\rm n}_{\rm O}\mu_{\rm O},\nonumber
\end{eqnarray}
where  E$_{\rm def}^Q$ is the total energy of the defected system in
charge state $Q$, $\mu_e$  the electron chemical potential (equaling
the Fermi level E$_{\rm F}$ in our T=0 calculations), and M$(Q)$  a
defect- and charge state-dependent multipole correction\cite{8,9}. The
${\rm n}_{\rm Si}$, ${\rm n}_{\rm Ge}$  and   ${\rm n}_{\rm O}$
 atoms considered in the modeling of a  specified defect,
possess the  chemical potentials $\mu_{\rm Si}$, $\mu_{\rm Ge}$, and
$\mu_{\rm O}$, discussed further below.   

Energies and forces are calculated from 
first-principles within   density-functional theory in the local
approximation, using the ultrasoft pseudopotential plane-wave method
as implemented in the  VASP code  \cite{9}. An isolated defect is
simulated in periodic boundary  conditions via repeated tetragonal
71-- and 72--atom  supercells of  crystalline $\alpha$-quartz SiO$_2$
having the (theoretical) linear dimensions 18.49, 
16.02, and 20.44 atomic units. Atomic geometries
 are optimized for all $Q$s (which are modeled by removing or adding
electrons as appropriate, the added charge being compensated by a 
uniform background) until all residual force components in the
system are below 0.01 eV/\AA. No symmetry restriction is imposed. A
 (222) Monkhorst-Pack mesh is used \cite{9}
for  k-space summation (4 points in the supercell  Brillouin
zone).  Total energy differences of different charge states define
ionization energies,  i.e. the energy  needed to promote e.g. an
electron from the valence band into an empty acceptor level. Formation
entropies are beyond the scope of the methods used here; plausible
estimates are used when needed. 

\begin{figure}[h]
\epsfclipon \epsfxsize=8cm \epsffile{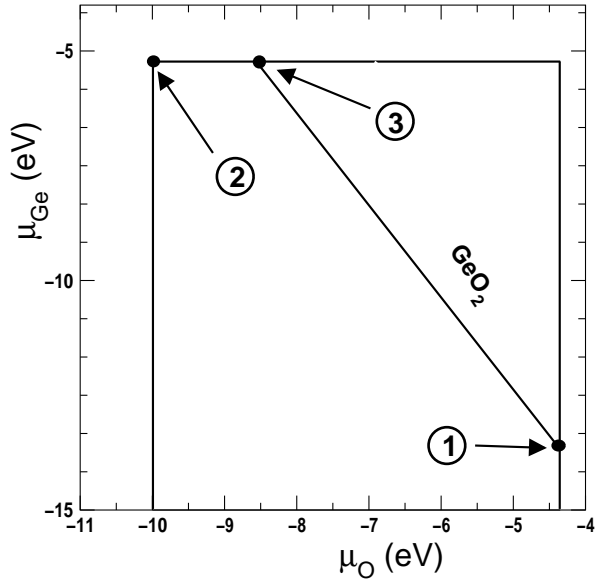}
\caption {Solubility region  of Si-substituting Ge in SiO$_2$
as a function of the impurity and oxygen chemical potentials.  
Relevant limiting values of the latter are indicated, numbered 
as in the text. The oblique line is the solubility limit due to  GeO$_2$
formation. The limit given  by GeO formation is irrelevant and is not
displayed.}
\label{fig1}
\end{figure}
To determine the formation energy of Ge and related defects, we must
discuss the  thermodynamical growth conditions of Ge-doped
SiO$_2$. The chemical   potentials fall within a range determined by
the  energy  $\mu^{\rm bulk}$ of the  condensed phase of each atom
(``bulk'' oxygen is assumed to be the oxygen molecule),  and the
calculated formation enthalpies $\Delta {\rm H}^{\rm SiO_2}$=--11.31 eV
 of SiO$_2$ and  $\Delta {\rm H}^{\rm GeO_2}$=--8.80 eV of GeO$_2$,  the 
most stable compounds liable to form out of O, Si, and Ge. The
following conditions  apply: 
\begin{eqnarray}
\mu_{\rm Si}^{\rm bulk}+\Delta{\rm H}^{\rm SiO_2} &\leq &\mu_{\rm Si} 
\leq \mu_{\rm Si}^{\rm bulk},\nonumber\\
\mu_{\rm O}^{\rm bulk}+ 0.5\, \Delta{\rm H}^{\rm SiO_2} &\leq &\mu_{\rm O} \leq
\mu_{\rm O}^{\rm bulk},  \nonumber\\  
\mu_{\rm Si}+2 \mu_{\rm O}=\mu_{\rm SiO_2}, &  &
\mu_{\rm Ge}+2 \mu_{\rm O}\leq\mu_{\rm GeO_2} \nonumber
\end{eqnarray}
The relevant extremal 
conditions are then   
\begin{enumerate}
\item O-rich, Si- and
Ge--lean :  $\mu_{\rm O} = \mu_{\rm O}^{\rm \rm bulk}$,
$\mu_{\rm Si}= \mu_{\rm Si}^{\rm bulk}+\Delta{\rm H}^{\rm SiO_2}$
and  $\mu_{\rm Ge}= \mu_{\rm Ge}^{\rm bulk}+\Delta{\rm H}^{\rm GeO_2}$;
\item Ge- and Si-rich, O-lean : $\mu_{\rm Ge}=\mu_{\rm Ge}^{\rm bulk}$,
$\mu_{\rm Si}=\mu_{\rm Si}^{\rm bulk}$, and $\mu_{\rm O}=
\mu_{\rm O}^{\rm bulk}+0.5\, \Delta{\rm H}^{\rm SiO_2}$; 
\item Ge-rich, O-lean with respect to GeO$_2$, and Si intermediate
 : $\mu_{\rm Ge}=\mu_{\rm Ge}^{\rm bulk}$, $\mu_{\rm Si}=\mu_{\rm
Si}^{\rm bulk}+ (\Delta{\rm H}^{\rm SiO_2}-\Delta{\rm H}^{\rm
GeO_2})$,  and $\mu_{\rm O}= \mu_{\rm O}^{\rm bulk}+0.5\, \Delta{\rm H}^{\rm
GeO_2}$.  
\end{enumerate}
These three cases are indicated in Fig. \ref{fig1}, which depicts the
solubility region of Ge in SiO$_2$ in the \{$\mu_{\rm
O}$,$\mu_{\rm Ge}$\} plane.

 To substitute as much   Si as possible with Ge, one should work in
comparatively Ge-rich and Si-lean conditions. Case 3 above, the most
favorable in  this respect,  demands at the same time that O be lean with
respect to GeO$_2$ formation. As a consequence, the material will
contain both a high Ge concentration, and a  large amount of oxygen
deficient centers (as  shown below, these tend to localize
near Ge atoms). The minimum formation energy of   Ge$_{\rm Si}$,  0.88
eV, is obtained indeed in case 3  above. The ensuing
solubility of Ge at a tipical  growth temperature T$_g$=1700 K (which
we assume throughout) is 0.2 mol \%. This
encouragingly high value falls well within the  experimental  range
assuming a  plausible formation entropy of 3--4 k$_{\rm B}$. 

\begin{figure}[h]
\epsfclipon
\epsfxsize=8cm
\epsffile{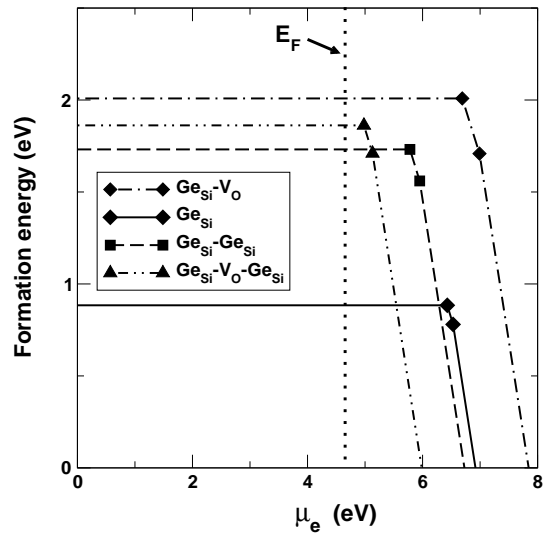}
\caption {Formation energy of  Ge-related defects: Ge$_{\rm Si}$,
Ge-adjacent oxygen vacancy, substitutional Ge pair with and without 
O vacancy. The slope of the formation energies is equal to the charge 
state (possible values are 0, --1, and --2 in the present case).
The vertical dotted line indicates the calculated Fermi level.
All the centers possess accessible paramagnetic ($Q$=--1) states.}
\label{fig2}
\end{figure}

Fig.\ref{fig2} shows the formation energies in various charge states, 
 as a function of the Fermi level and
 for  the growth conditions discussed above,
of all the centers considered here, and discussed in turn below:
 substitutional Ge$_{\rm Si}$, the  oxygen vacancy
V$_{\rm O}$ adjacent to Ge$_{\rm Si}$,
the Ge$_{\rm Si}$-Ge$_{\rm Si}$ pair (substituting the
 central Si of two neighboring tetrahedra),
the Ge$_{\rm Si}$-V$_{\rm O}$-Ge$_{\rm Si}$ complex (the oxygen
bridging the Ge-Ge pair is missing).
The dotted vertical line marks the Fermi level as calculated in 
the presence of all the various defects, 
namely E$_{\rm F}$ = E$_v$ + 4.65 eV at room temperature. 
We assumed a gap of 9 eV, and the band edge effective masses 
$m_v\simeq$10 $m_e$ and $m_c \simeq$0.3 $m_e$ as  estimated  in
 Ref. \cite{mstar-sio2}.   Indeed, all the  defects considered are
in their neutral state at room temperature.

Ge$_{\rm Si}$ is neutral over most of the Fermi level range,
and captures electrons in strong $n-$type conditions.
 Structurally, neutral Ge$_{\rm Si}$  is relatively
trivial, with Ge-O bonds  (unsurprisingly slightly longer, 1.67 and
1.68 \AA, than the bulk Si-O bonds of 1.58 and 1.59 \AA) and the other
structural parameters (Ge-Si distance : 3.09 \AA, O--Ge--O angle :
109$^{\circ}$, and Ge--O--Si angle : 142$^{\circ}$)  being
characteristic of an isotropically expanded but otherwise regular
tetrahedron.       
 The charge state  $Q=$--1, one of the models proposed
\cite{Pacch} for the paramagnetic   Ge(1)  center, is not
stable in the as-grown material, but it may be observed in
 $n$-type or  electron-irradiated material. 
Upon electron capture, the Ge substitutional moves off-center
causing an orthorombic distortion with two short  
and two long Ge-O bonds (differing by $\sim$ 0.2 \AA), in good
agreement with the report of Ref.\cite{Pacch}.

The properties of Ge$_{\rm Si}$ also match the main Ge-related  optical
signature in highly doped samples, namely the apparent reduction of
the  absorption gap to  about 7 eV \cite{carb}. Indeed,  our calculated
concentration and electronic structure  predict an absorption  into
the Ge$_{\rm Si}$ extrinsic level starting around 6.5-6.8 eV with an 
effective final density of states (DOS) of about $\sim 10^{20}$  cm$^{-3}$.
Assuming an  impurity band width of 1 eV, likely to occur
at such high Ge concentrations, it appears that the 
valence-to-Ge$_{\rm Si}$ extrinsic state is as  relevant as the
 fundamental interband transition, even accounting for reduced
 oscillator strength, since the effective conduction band
DOS of SiO$_2$ is in fact about  1$\times$10$^{19}$  cm$^{-3}$ at room 
temperature.  
In summary, the large absorption red shift in Ge:SiO$_2$
 is impurity-related, as opposed to a standard alloying effect; 
a simple Si$_{1-x}$Ge$_x$O$_2$ alloying picture 
would predict (the gap of GeO$_2$ being 5.6 eV) 
a shift of less than 0.1 eV instead of the 
observed $\sim 2$ eV at the  typical 
Ge concentration $x\sim 0.02$. The behavior just discussed is
 similar to the  impurity-level (or -band) effects observed 
e.g. in  InGaAsN at  low N concentration \cite{ingaasn}.  

We now turn to V$_{\rm O}$-Ge centers.
The isolated oxygen vacancy at the chemical potentials giving
 maximum Ge incorporation,  has a formation energy of 3.31 eV,
i.e. 0.4 eV less than in pure SiO$_2$ in stoichiometric conditions.
At  T$_g$=1700 K, this implies a concentration increase of a mere
factor  of 10. A vacancy  in the same simulation cell, but as far as
possible from  Ge$_{\rm Si}$ (about  7 \AA\, away) has the same
formation energy, i.e. it is effectively decoupled from the Ge
substitution.  However,  when the vacancy and Ge$_{\rm Si}$ are first
neighbors, the formation energy drops to 2.0 eV   (Fig. \ref{fig2}),
implying a large attraction for vacancies towards Ge$_{\rm Si}$
centers. Now {\it this} reduced formation energy implies that the
concentrations of  Ge-coupled vacancies 
is about 10$^{4}$  that  of distant vacancies, and 10$^{5}$ 
that of those in pure SiO$_2$, in quantitative agreement with the
observed ratio  \cite{Hoso}.  Therefore the growth  conditions of
Ge-doped SiO$_2$ naturally produce an off-stoichiometry of
oxygen: Ge-doped SiO$_2$ contains of the order of 10$^5$ more
``simple'' oxygen deficient centers than  pure SiO$_2$ does. 
In addition this oxygen deficiency is Ge-biased: almost all 
of the vacancies are localized near Ge$_{\rm Si}$ sites, due to
the  attractive Ge$_{\rm Si}$-vacancy effective interaction, and to 
a lesser extent to the different solubility limits  provided 
by  GeO$_2$ and SiO$_2$. This applies also, as discussed further
below, to the vacancy bridging a pair of Si-substituting Ge.

The  ground state of the  Ge-neighboring vacancy
V$_{\rm  O}$--Ge$_{\rm Si}$ is neutral and unpuckered -- i.e. the
cations neighboring the vacancy remain near the vacant site, and do
not undergo distortions of the type involved in the E$'$ center
\cite{noi}. It captures electrons only for extreme $n$ conditions.
 The {\it  unpuckered} +1 charge state  is never
thermodynamically stable, which  rules out, e.g.,  the possibility of
associating  this defect to  the Ge(0) or  Ge(1)  centers
\cite{FGS,Pal}. Instead,  similarly to the undoped case
\cite{noi}, V$_{\rm O}^+$   becomes a  metastable ground state in a
puckered configuration (lower than the  unpuckered one by  0.15 eV)
if the Fermi level is below about 1.6 eV \cite{nota}.
 Thereby, it may be a candidate
for the Ge(3) center \cite{FGS,Uchi,Fuji}.  The  E$'$-like off-site 
puckering involves  the Si atom  adjacent to the  vacancy, since  
the puckering of Ge (on the  correct  side of the vacancy \cite{noi})
costs 1.3 eV more than that  of  Si.  Consequently, the Ge$_{\rm Si}$
is left  with an  EPR-detectable unpaired electron, in agreement with
the fact that the EPR lines of Ge(3)  show\cite{epr-ge} the
characteristic  signatures  of hyperfine interaction with $^{73}$Ge.

Given the larger amount of vacancies and their preference for the
 vicinities of Ge, and accounting for  the $^{73}$Ge and
$^{29}$Si   isotopic abundances of $\sim$8\% and 4\%, the
concentration of Si-Ge E$'$ measured by EPR should  be a factor of
10$^4$ that of E$'$ in SiO$_2$.  Observed values
\cite{crivelli} are in a range upwards of 10$^2$. 
 In  agreement with the observed axial  simmetry   of Ge(3), 
the bond lengths between the EPR-active Ge and the 
three first-neighbor oxygens are about  the same.
 Also, the fact that the puckered
configuration is lower in energy than the unpuckered one agrees with
the observation of Ge(3) defects even in non irradiated  samples
\cite{FGS}.   

As mentioned, the thermodynamical stability of Si-Ge E$'$ is possible 
only at rather extreme $p$ conditions ($\mu_e\sim$1.6 eV),
 not realized in as-grown material. This center should therefore
be observed only upon (radiation-induced or electrical) hole injection.
In as-grown material, it may still be possible that, after its
excitation to the positive   state by e.g. an optical excitation, the
return of Ge-Si  E$'$ to the  neutral (and  thence to the unpuckered)
state is slowed down due to selection  rules and/or disorder, neither of
which have been considered here.    

So far we identified possible candidates for the 
 Ge(1) electron-capturing and Ge(3) E$'$-like centers. We now
 move on to  Ge-pair defects, consisting of   two corner-sharing  
Ge-centered tetrahedra. In the neutral state, the tetrahedra exhibit 
Ge-O bonds and angles close to those of the isolated Ge$_{\rm  Si}$ 
in their neutral state. 

As seen in Fig. \ref{fig2}, the formation energy of two  Ge impurities
at neighboring substitutional sites is almost exactly twice that of isolated 
Ge$_{\rm Si}$; therefore, there is neither a driving force for, nor
an  energetic hindrance against, the clustering of Ge$_{\rm Si}$. Kinetics,
and hence thermal and growth history will play the deciding role. 
Once more, the $Q$=--1  charge  state is not a
ground state of the defect in as-grown material, but it may be 
observed in $n$-doped irradiated material, and thereby become a
candidate   for the Ge(2) paramagnetic center, as suggested in 
Ref. \cite{kawazoe}. 

Finally we consider  the formation of V$_{\rm O}$ between the two Ge atoms:
it costs 0.15 eV less than between a Ge and a Si (Fig. \ref{fig2}), 
because of the previously discussed natural tendency of the two Ge to 
host off-stoichiometry of oxygen nearby them; this gives a concentration 
of a further factor of 3  higher than for the Ge-neighboring vacancy.
In this case, the negative charging level is just slightly  ($\sim$ 0.1 eV)
above the Fermi level so that, given the uncertainties on the 
energy levels, it may well turn out to be already occupied
in weakly, unintentionally 
$n-$type as-grown material, or will readily become occupied in moderately
$n-$doped or irradiated material. Thus Ge-V$_{\rm O}$-Ge
  stands out as well as a candidate   for the Ge(2) paramagnetic 
center\cite{FGS}.  

Once more, the  unpuckered +1 vacancy is 
not a stable  ground state;  its  E$'-$like
puckered-Ge configuration becomes metastable below $\mu_e$=1.5 eV 
\cite{nota}.
As the  metastable energy minimum is found to be 0.45  eV higher than
the undistorted one, the EPR-active Ge-pair E$'$--like configuration  is
far less frequent than the Si-Ge one
($\sim$10$^{-7}$ at room temperature, although it is 
possible that, as for the standard E$'$,
 a  confinement barrier  exists prolonging its existence).
 This suggests that the proposed  \cite{Hoso}
attribution of the Ge(3) center to this complex is unlikely to be
correct.

In summary, we have shown that thermodynamical growth conditions of
Ge-doped SiO$_2$ naturally produce off-stoichiometry of oxygen, and that
oxygen deficient centers form preferentially near Ge impurities. 
We find that energetics  is neutral as to clustering of substitutional
Ge, so that kinetics will be important.
The calculated concentration (in agreement with experiment)
and electronic structure of Ge$_{\rm Si}$ show a large density of
states starting at 6.5-6.8 eV which may explain the apparent gap reduction
to about 7 eV in  absorption. Our results contribute to the
identification of  the  Ge(1), Ge(2), and Ge(3) centers. Ge(1) and Ge(2)
 are electron-capturing paramagnetic centers; the former is probably related
 to the singly negative states of Ge$_{\rm Si}$; the latter may be associated 
with the singly-negative Ge$_{\rm Si}$-Ge$_{\rm Si}$ or
Ge$_{\rm Si}$-V$_{\rm O}$-Ge$_{\rm Si}$ complexes. Ge(3) may instead
 be attributed to the Ge-Si E$'$-like 
center. Hyperfine-parameter calculations are underway to  pinpoint
these attributions. In any case, these centers should be 
preferentially observed  in $n$-doped or irradiated material.   

Work supported in part by the Parallel Supercomputing
Initiative of INFM, and the European Union within the INVEST project.

\end{multicols}

\begin{thebibliography}{99}
\bibitem{nota} At this energy,
  the puckered +1 state drops below the puckered neutral;
the  unpuckered neutral is always lower than unpuckered +1 (see 
e.g. \cite{noi} for more details, in particular Fig.2).
\end{thebibliography}
\end{document}